# COLLECTIVE CREATIVITY: WHERE WE ARE AND WHERE WE MIGHT GO

| Lixiu Yu | Jeffrey V. Nickerson | Yasuaki Sakamoto |
|---|---|---|
| Carnegie Mellon University | Stevens Institute of Technology | Stevens Institute of Technology |
| Pittsburgh PA 15213 | Hoboken NJ 07030 | Hoboken NJ 07030 |
| lixiuyu@cs.cmu.edu | jnickerson@stevens.edu | yasuaki.sakamoto@stevens.edu |

**ABSTRACT**

Creativity is individual, and it is social. The social aspects of creativity have become of increasing interest as systems have emerged that mobilize large numbers of people to engage in creative tasks. We examine research related to collective intelligence and differentiate work on collective creativity from other collective activities by analyzing systems with respect to the tasks that are performed and the outputs that result. Three types of systems are discussed: games, contests and networks. We conclude by suggesting how systems that generate collective creativity can be improved and how new systems might be constructed.

**INTRODUCTION**

Even though we admire individual inventors, most creative output is a result of social processes involving many (Amabile 1996; Fisher 2004; 2005). Until recently there were two levels to be taken into account: individual creativity and group creativity. But the advent of collective intelligence has changed things, adding a category qualitatively and quantitatively different from either level: Work can be delegated not to a group but to a set of strangers assembled to perform a task, a crowd (Howe 2008; Malone 2009; Quinn and Bederson 2011). The crowd has shown its power in solving many problems, this power has attracted attention from both the academy and industry.

The type of work produced by the crowd widely ranges from image labeling, information collection, scientific problem solving to idea generation (Von Ahn and Dabbish 2004; Kittur and Kraut 2008; Cooper et al. 2010; Yu and Nickerson 2011; in press). All this work can be classified into different categories according to different criteria. In this paper, we focus on the tasks performed by the crowd and the end output of the collective activities. We distinguish the crowd's creative work from other collective activities and define collective creativity based on this distinction.

Research on the use of crowds to perform complex creative work is difficult because of the nature of the tasks and the nature of the crowd. Creative tasks often lack evident solutions: There are usually no black and white answers, and judging criteria can be difficult to agree on. The work arguably requires a high degree of cognitive effort from each member of the crowd. At the same time, members of the crowd often lack expertise relevant to the task. Moreover, the crowd is physically distributed and is hard to organize and engage in dynamic interaction. Consequently, collective creativity is still in its early stages.

Nevertheless, the number of new systems intended to generate collective creativity is beginning to grow rapidly. These systems occur in different domains using diverse methods. Here we paint the landscape of collective creativity research by reviewing several systems in which crowds have been used to perform creative work.

We proceed by defining collective creativity in terms of a system's tasks and outputs. Then, we classify current systems. Finally, we suggest ways to improve existing systems and create new ones, focusing on the possibility of creating reflective systems that can improve both the experience of the crowd and the creativity of the system's output.

**DEFINING COLLECTIVE CREATIVITY**

In defining collective creativity, we identified two dimensions that collective activities vary on, shown in Table 1: the nature of the task, and the nature of the output of the collective activity.

These two dimensions respect the definitions of creativity itself. That is, some researchers focus on the creativity of the process through which products are produced (Wallas 1926), and others focus on the creativity of the output (Kozbelt, Beghetto, and Runco 2010). Collective creativity is about both the creativity of the output from a collective process and the creativity of the tasks performed as part of this process.

Table 1. Dimensions of Collective Activities

|  | **Routine tasks** | **Non-routine tasks** |
|---|---|---|
| **Emergent output** | Galaxy Zoo<br>Photocity<br>Picbreeder<br>Manatee | **Collective creativity**<br>Foldit, Eterna<br>Threadless<br>Innocentive<br>Quirky<br>Climate Colab<br>All Our Ideas<br>CrowdForge<br>Evolutionary Collective Design<br>Polymath |
| **Aggregated output** | Soylent<br>ESP<br>reCaptcha<br>Crowdfunding<br>Tiramisu | Wikipedia |

We begin by looking at the variety of tasks performed by the crowd. Some tasks are routine: the tasks can be done more or less automatically with little uncertainty about one's ability to accomplish the task in a set amount of time. Tasks in this category include transcribing spoken language to text, translating from one language to another, and labeling images. While each involves human cognition, given time, it is likely that the task can be completed. Examples of these activities are shown under "Routine tasks" in Table 1.

By contrast, other tasks entail great uncertainty about the ability of the performer to find a solution in a given amount of time. For example, a large search space leaves open the possibility that a person may find a result early, due to blind luck, or after a considerable amount of time, or never. This category of task requires more cognitive effort and its exploration may lead to solutions that have not been found before: such solutions and the exploration process itself are creative. Within this category there are two subcategories. First, tasks for which there are simple computational ways to evaluate the output. For example, the discovery of a new model for folding a protein can be confirmed through a simple calculation. Still other tasks are more open ended, in the sense that the evaluation of generated solutions cannot be done computationally. For example, the design of an everyday artifact – a chair, an alarm clock or a door – would likely need to be judged along several amorphous dimensions – such as comfort, convenience, and aesthetics – and these dimensions can change depending on the objectives of the design, as well as the context for use. In sum, at a high level we can categorize tasks into routine tasks that require less cognitive effort and non-routine tasks that require more cognitive effort.

Next, we look at the nature of the final output of the collective activity. Some activities' outputs are more or less collections of each individual's work, which we call *aggregated output*. Some activities will produce outputs that are new: the outputs are *emergent*. There are two ways new things can emerge. First, through discovery: crowd participants can find something not seen before. Second, through combination: integrating participant's work can produce something novel.

For example, a system that asks people to record when they step onto a bus will produce a useful schedule, an aggregation of times (Zimmerman et al. 2011); a system that asks people to classify telescope images as containing quasars might lead to the discovery of something new (Lintott et al. 2008).

In Table 1, we classified samples of collective intelligence systems from different fields according to the two dimensions: the nature of the individual task, and the nature of the final output. The samples were selected based on their mention in conference proceedings and journal papers. The samples represent a wide range of collective activities.

Wikipedia in this scheme is the result of the non-routine task of article writing, but the output is predictable, a result of the tight specifications enforced by the editors. In particular, Wikipedia rules prohibit the introduction of new research: instead the articles are expected to summarize, to aggregate existing knowledge.

In the lower left quadrant, Soylent is an example of a system through which crowd workers proofread a paper in real time (Bernstein et al. 2010). The work preformed is routine, and the end output is a collection of the copyedited sentences. Nothing substantially new is meant to emerge from this process. Similarly, the ESP game produces labeled images as a result of crowd participation (Von Ahn and Dabbish 2004); the output is an aggregation of labels. Recaptcha helps digitize books when the crowd solves puzzles, and the end output is a collection of readable words (Von Ahn et al. 2008). Crowdfunding results in a collective financial support when people pool their money together to support a cause (Chuang and Gerber 2011).

There is a class of applications that involve routine work, but can generate novel results. For example, Galaxy Zoo is the result of many people classifying images of the stars: their work can, and has discovered new astronomical phenomena (Lintott et al. 2008). The project called Photocity sent a crowd out to take pictures of buildings, and out of this routine task a detailed three-dimensional model emerged (Tuite et al. 2011). In Picbreeder, people simply judge computer-generated images, and through an evolutionary process, complex images emerge (Secretan et al. 2008). Manatee asks people to come up with a subject; the system then uses search engine technology to text mine recent discussions of the subject, generating an original comic strip (McNally and Hammond 2011).

The examples of crowd activities discussed so far are societally useful, and these systems require a great deal of creativity on the part of the designers. They are harbingers of future collective creativity systems. Nevertheless, we do not consider these systems as central instances of collective creativity, because in many cases, the creative output is as much due to the cleverness of the system's programmers as to the contributions of the crowd.

Instead, we focus on the examples in the upper-right quadrant: the crowd performs non-routine tasks and the end output is emergent. That is, something new is discovered, or something new emerges from the sum of the parts. With respect to task type and output type, we define a collective creativity system as one in which crowds engage in non-routine tasks through which novel output emerges.

## TYPES OF COLLECTIVE CREATIVITY SYSTEMS

### Overview

Looking at the set of systems in the top right quadrant, we can see they fall roughly into three categories. First, there are games: Foldit and EteRNA. Then, there are contests: Threadless, InnoCentive, Climate CoLab and Quirky. Third, are systems with more elaborate structures based on networks: CrowdForge and Evolutionary Collective Design.

### Games

Games have been built to aid scientific discovery. For example, players create protein structure models by playing an online game called Foldit (Cooper et al. 2010). The results show that for difficult problems, the players' solutions can in some cases outperform computational methods. The game creators were interested in finding out how the participants were discovering new models, so they made available scripts and script editors, to encourage the participants to codify and share their strategies (Khatib et al. 2011). These novices discovered an algorithm at the same time a scientific team did. Moreover, the crowd's algorithm was better optimized for the game environment.

Another related game is called EteRNA. It allows participants to design RNA with the help of automated tools (EteRNA 2011). Particularly novel solutions are then tested in wet labs to verify the models.

Thus, the crowd in games can perform high demanding complex tasks and the output of the games is the solution to difficult scientific problems. Moreover, this collective creative process can be traced, providing us insight on how collective creativity operates.

### Contests

Online contests, and more specifically idea competitions, are used for eliciting collective creativity. Challenges are posted with prizes promised, and the crowd competes with each other for the prizes by providing solutions. The hosts of idea competitions are various including companies, government institutions, and third party websites. The tasks cover a wide range of domains, including scientific problems and product designs. The prizes come in many forms: some are monetary and others are reputational. For example, one of the successful idea competition websites is Threadless (Brabham 2008). Threadless makes open calls for the design of T-shirts on a particular theme periodically. The crowd submits their designs, and the best ones are selected to be manufactured and sold. The same contest form is also applied by InnoCentive (Brabham 2008) and Climate Colab (Malone and Klein 2007) with minor variations. The main difference is that the challenges of these two later sites involve scientific problems.

Another contest with a slightly different process is Quirky (Maher 2010). Participants submit ideas, and pay for the privilege to submit. Other participants make suggestions, and thereby earn influence points. Those that achieve some threshold of interest from the crowd are directed to the site owners, who decide which designs to put into test production. Since many of the items can be built on demand through outsourced manufacturing (or 3D printers), the designs are posted, the idea publicized, and once someone wants one, an item is produced.

Ideally, every individual and organization would be able to hold their own contests for their own problems. Tools to support such broad participation are emerging. For example, the survey tool "All Our Ideas" allows users to quickly set up a free website where large number of people can contribute and rank ideas (Emery 2010).

While games and contests share some attributes – for example the use of rankings as an incentive – on others they differ dramatically. In particular, game play provides rich and continuous feedback. Contests do not.

This advantage of games comes with a cost: the development of a game is complex and expensive, whereas the announcement of a contest can be accomplished without a large investment in technology. However, contests often involve legal complexities: for example, InnoCentive provides an intricate set of rules relating to the sharing of prize money. Games, in this respect, are often simpler. Indeed, people engage in both Foldit and Eterna without being compensated monetarily.

**Networks**

Both contests and games have fairly simple organizational structures. People either compete as individuals, or as teams. But in other kinds of collective creative work, more intricate structures are used: we will call this category *networks*. We highlight two examples here.

In the previous two categories, each member of the crowd is almost interchangeable: they can more or less direct their own tasking, proceeding at their own pace. The interdependence of the crowd is low. By contrast, in CrowdForge, a problem is subdivided, each member of the crowd gets his or her own piece to work on, and then the pieces are aggregated later (Kittur, Smus, and Kraut 2011). This framework was used to produce articles, and also to research purchasing decisions.

The concept behind this system is related to two computer techniques: divide and conquer, in which problems are broken down into smaller bits and then solved (Horowitz 1977; McDonald 2011), and, more recently, map/reduce, in which ideas are parallelized and then aggregated (Dean and Ghemawat 2008). In this way, CrowdForge is consistent with the overall theme of human computation (Von Ahn 2006): a problem is defined in a way that could be performed by computing nodes, and humans are substituted for these nodes.

That such networked crowdsourced processes work at all is intriguing. Hierarchical specialization is a well-understood part of business management, but such specialization usually relies on stable relationships and organizational learning. Collective activities based on a map/reduce structure have neither of these social affordances, yet the output produced can be useful.

Map/reduce is not the only paradigm that can be implemented by the crowd: Evolutionary algorithms (Goldberg 1989) can also be implemented In one instantiation of this idea, a computer generates new alternatives, and humans pick the fittest to evolve further (Secretan et al. 2008). In a more radical scheme, the crowd both generates and evaluates designs (Nickerson and Sakamoto 2010; Yu and Nickerson 2011; in press). This structure has been applied to several design problems and shown to be effective in producing creative ideas.

The algorithms here function as organization structures that coordinate interaction. Crowd members only communicate through the designs they produce. The first crowd generates ideas, the second crowd evaluates ideas and the third crowd generates new ideas by selecting and combining existing ideas. Creative ideas emerge through this evolutionary process.

Both of these examples involve intricate and predefined structures. It is also possible for networks of interaction to emerge organically. For example in Polymath (Cranshaw and Kittur 2011), a set of mathematicians, professionals and amateurs, set out to make mathematical discoveries together. What emerged was not only new proofs, but also a way of working in which these proofs were annotated, hyperlinked, and blogged about in a distributed manner. Professionals and amateurs played different roles, but amateurs were able to contribute to mathematical construction of the proofs.

There are examples that do not fit as obviously into these categories. In particular, SwarmSketch (2011) invites participants to draw a simulated ink line on a collective portrait, and then vote for the lines of previous participants. This is neither a contest nor a game. But there is a network element: the participant is joining a network of previous participants, and then affecting the work by contributing a line and either strengthening or weakening other lines.

The types of these collective creativity systems are summarized in Table 2.

**Table 2. Types of collective creativity**

| Category | Examples |
|---|---|
| Games | Foldit  Eterna |
| Contests | Threadless  InoCentive  Climate CoLab  Quirky  All Our Ideas |
| Networks | CrowdForge  Evolutionary Collective Design  Polymath  SwarmSketch |

**FUTURE DIRECTIONS**

In uncharted territory, there are many possible paths for exploration. Here, we suggest areas to focus on in future research. We start with the three types of systems just described in the previous section – games, contests, networks – and then discuss the possibility of creating reflective systems that can improve both the creative outputs and the experience of the crowd.

**Games**

Using online games to generate collective creativity is still in its early stage; currently, computer games are mainly used for scientific problem solving. But there are other domains game techniques might be useful in. In particular, game techniques might be applied to design problems; the game techniques might be an effective way to engage the crowd in working on open-ended problems. For example, the prototype architecture design of a new concert hall might be realized through a game with parameters that can be altered – materials, façade, structure – as well as evaluation functions which simulate acoustics.

**Contests**

Online contests are favored commercial strategies to harness crowd creativity. These contests are designed for a wide range of topics and have been demonstrated to be effective. However, rewards don't necessarily improve the quality of the submitted ideas (Walter and Back 2011). And contests can be a waste of the crowd's time and effort because so few entries end up being used. Research on group idea generation shows ideas improved if they were built on other ideas (Kohn, Paulus, and Choi 2011). Therefore, competitions that permit idea reuse, such as those hosted by Matlab (Gulley 2001), might produce better outcomes. One interesting example of reuse is Scratch: youths have created their own contests on this website, giving each other rewards for creating interesting drawings, animations, and computer games (Nickerson and Monroy-Hernandez 2011). As part of these contests they remix each other's work. These examples suggest the creation of collective creativity environments in which the contests serve to motivate, but pure competition is relaxed in favor of encouraging participants to build on the contributions of others.

**Networks**

Most studies in the networks category were conducted in crowdsourcing marketplaces such as Amazon Mechanical Turk (Kittur, Chi, and Suh 2008). Workers finish short sub-tasks in return for small payments. For complex tasks, such as open-ended writing or design, one of main impediments to successful crowdsourcing is the difficulty of building and maintaining the crowd's motivation; indeed, economic forces interact with motivation to determine the the amount of compensation the crowd requires to perform difficult tasks (Toomim 2011). For simple tasks, such as image ordering or word puzzles, financial incentives appear to influence the quantity, but not the quality, of the work performed (Mason and Watts 2009). In games and contests, the incentives that keep the crowd motivated include the ranking of winners, the fun of playing, instant rich feedback and prizes. Future systems that structure the crowd into networks might be more effective if they incorporate such incentives (Dow et al. 2012; Luther, Diakipoulos, and Bruckman 2010). Moreover, given a network structure, incentives to engage and collaborate might be distributed through network links, in an approach similar to one used in solving search problems (Pickard et al. 2010).

**A Step Toward Reflective Systems**

We differentiated between routine tasks, which take small amounts of cognitive effort, and non-routine tasks, which take more effort, and are potentially more frustrating. Over time, such non-routine tasks can become more and more automatic: this is the process of acquiring expertise through reflective practice (Ericsson 2005; Simon and Chase 1973).

Such a process is important because it allows us to make common tasks more routine so that we can focus our cognitive effort on higher-level tasks. From the standpoint of collective creativity systems, it is desirable if tasks that are routine can be automated so that people's attention can be devoted to more complex activities: the system then becomes more powerful. We see examples of this in some game-based systems. For example, in EteRNA (2011), players are presented with simple molecules to build, and once such tasks become automatic they are presented with problems neither computers nor experts alike can solve in routine ways. In Foldit (Khatib et al. 2011), the participants were given the ability to script their actions, thereby automating their strategies. This not only allowed for the analysis of learning strategies of the participants, but also freed participants up to engage in more complex challenges. Specifically, they focused on creating new strategies for work rather than repeating existing strategies.

This suggests a promising direction: the creation of collective creativity systems that seek to foster learning. The learning can happen at four levels. First, the human participants learn to take complex cognitive tasks and make them routine. Second the systems can promote social learning, in which human participants learn from the output of others. Third, the computers can automate these routine functions using human performance as input to machine learning techniques, in the end freeing up humans to address more and more complex cognitive tasks. Fourth, learning at the three levels mentioned above results in learning at the community level, as both human and computer capabilities grow.

Such systems contribute to the society not only by generating creative solutions to social and scientific problems but also by building tools that augment human cognition and promote intellectual growth.

Designers of such systems should address an important ethical issue: what do participants in collective creativity gain? Ownership rules are murky: for example, a product may be designed by thousands of people from around the world, but legal systems cannot currently accommodate so many claims to ownership. Participants may or may not be paid, according to the crowdsourcing model. But participants might, regardless of other compensation, learn. Properly constructed collective creativity environments can help build expertise that may be valuable not just in the context of the system, but also in other endeavors. For example, those participating in Foldit and EteRNA are learning chemistry, biology, and computer science; those participating in Polymath are learning math and interacting with accomplished mathematicians. Consequently, system designers might consider how the needs of the system – to build expertise and focus cognitive activity – might coincide with the desires of participants – to learn and grow intellectually.

**CONCLUSIONS**

In this paper, we define collective creativity based on the nature of the task performed by the crowd and the nature of the output of the collective activity. Collective creativity involves non-routine tasks out of which new ideas emerge. We analyzed three different types of collective creativity systems: games, contests and networks. We also pointed out several future directions: how systems that generate collective creativity can be improved and how new systems might be constructed.

This review suggests the design space for collective creativity is large, and mainly unexplored. By combining features of these different systems, we might discover new systems. For example, the interactive feedback of a game might be combined with the network organization of the Evolutionary Collective Design System to produce large ad hoc organizations working to produce new and useful products.

In the background, of course, is the question of who designs the systems, and which domains are explored. Over time, the crowd itself may have something to say about this: members might want to contribute to their own destiny. Indeed, there is a feedback loop. Crowds configured in certain ways may achieve results that change the economic landscape, affecting the structures they work within. Then, a collective creativity system becomes a co-creation between the founders of the system and the crowd members, who both design the products and contribute to the design of the system. Out of this comes a kind of system that benefits society through its creative outputs and benefits participants through the mastery they attain.

## ACKNOWLEDGEMENTS

Funding for this research was provided by the National Science Foundation, awards IIS-0855995 and IIS-0968561.